\documentclass[aps,prl,superscriptaddress,longbibliography,twocolumn]{revtex4-1}

\usepackage{lmodern}

\usepackage[latin9]{inputenc}
\usepackage{amsmath}
\usepackage{amssymb}
\usepackage{graphicx}
\usepackage{epstopdf}
\usepackage{color}
\usepackage{enumerate}
\usepackage{ulem}

\begin{document}

\title{Joule overheating poisons the fractional ac Josephson effect in topological Josephson junctions}

\author{K\'{e}vin Le Calvez}
\author{Louis Veyrat}
\author{Fr\'{e}d\'{e}ric Gay}
\author{Philippe Plaindoux}
\author{Clemens B. Winkelmann}
\author{Herv\'{e} Courtois}
\author{Benjamin Sac\'{e}p\'{e}}
\email{benjamin.sacepe@neel.cnrs.fr}
\affiliation{Univ. Grenoble Alpes, CNRS, Grenoble INP, Institut N\'{e}el, 38000 Grenoble, France}

\date{\today}

\begin{abstract}
Topological Josephson junctions designed on the surface of a 3D-topological insulator (TI) harbor Majorana bound states (MBS's) among a continuum of conventional Andreev bound states. The distinct feature of these MBS's lies in the $4\pi$-periodicity of their energy-phase relation that yields a fractional ac Josephson effect and a suppression of odd Shapiro steps under $r\!f$ irradiation. Yet, recent experiments showed that a few, or only the first, odd Shapiro steps are missing, casting doubts on the interpretation. Here, we show that Josephson junctions tailored on the large bandgap 3D TI Bi$_2$Se$_3$ exhibit a fractional ac Josephson effect acting on the first Shapiro step only. With a modified resistively shunted junction model, we demonstrate that the resilience of higher order odd Shapiro steps can be accounted for by thermal poisoning driven by Joule overheating. Furthermore, we uncover a residual supercurrent at the nodes between Shapiro lobes, which provides a direct and novel signature of the current carried by the MBS's. Our findings showcase the crucial role of thermal effects in topological Josephson junctions and lend support to the Majorana origin of the partial suppression of odd Shapiro steps.
\end{abstract}

\maketitle

Topological superconductivity engineered by coupling superconducting electrodes to topological states of matter has attracted considerable attention due to the prospect of manipulating Majorana states for topological quantum computing~\cite{Kitaev03,Nayak08,Fu08}. Intense experimental efforts have focused on spectroscopic signatures of Majorana bound states (MBS's) in various superconductivity-proximitized systems, including semiconducting nanowires~\cite{Mourik12,Das12,Albrecht16,Deng16,zhang2017quantized}, atomic chains~\cite{Nadj-perge14,Jeon17} or islands~\cite{Menard17} of  magnetic atoms, and vortices at the surface of 3D TI's~\cite{Xu15}. 

\begin{figure}[h!]
	\includegraphics[width=1\linewidth]{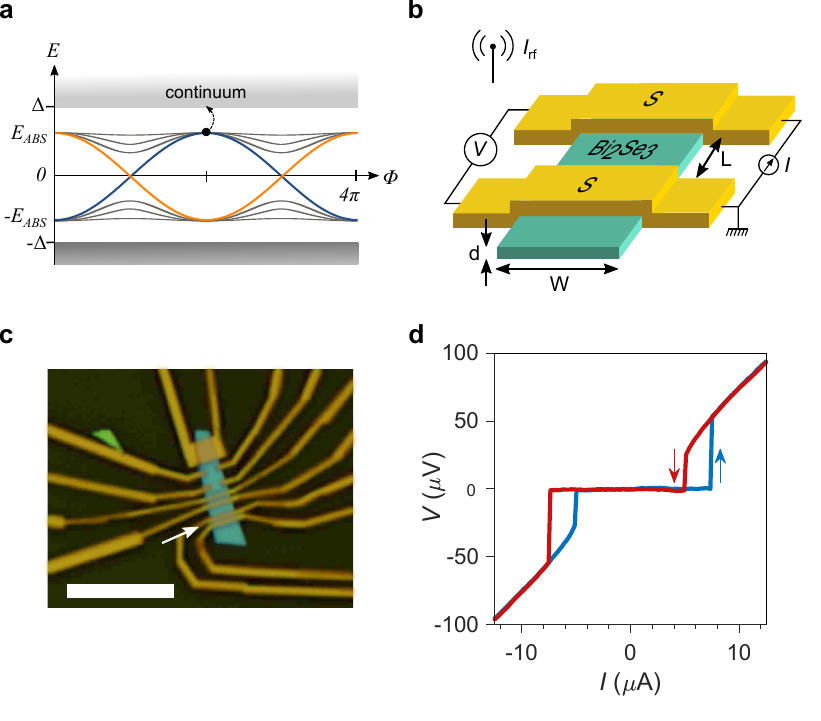} 
	\caption{\textbf{Topological Josephson junction on $\mathbf{Bi_2Se_3}$.} \textbf{a,} Energy-phase spectrum of the Andreev bound states for a topological Josephson junction at the surface of a 3D TI. $4\pi$-periodic, spinless, MBS's coexist with conventional $2\pi$-periodic ABS's. The maximum energy $E_{ABS}$ is lower than the quasiparticle continuum at $\Delta $ in case of imperfect interface transparency or the presence of a magnetic barrier~\cite{Fu08,Fu09,Snelder13}. \textbf{b,} Schematic of the Josephson junction geometry showing the superconducting electrodes (S) in orange that contact the top of the Bi$_2$Se$_3$ flake (in blue). \textbf{c,} Optical image of  device. Scale bar is $9\,\mu$m. \textbf{d,}	Current-voltage characteristic of the junction indicated by the arrow in \textbf{c}. Measurements were carried out at $0.05$~K. }
\label{Fig1}
\end{figure}

Another key approach to substantiate the very existence of MBS relies on the fractional ac Josephson effect~\cite{Kitaev01,Kwon04,Fu09} that develops in topological Josephson junction~\cite{Fu08,Fu09}. Theory predicts that MBS's shall emerge in such junctions as a peculiar, spinless Andreev bound state (ABS). Contrary to the conventional ABS's whose energy level varies $2\pi$-periodically with the phase difference $\phi$ between the junction electrodes, the MBS is $4\pi$-periodic and crosses zero-energy for a phase $\pi$  (see Fig. \ref{Fig1}a), yielding a fractional ac Josephson effect at frequency $f_J/2=eV/h$ ($e$ is the electron charge, $V$ the voltage drop across the junction and $h$ the Planck constant), that is, half the Josephson frequency $f_J$~\cite{Fu08,Fu09}.

Yet, revealing such a $4\pi$-periodic contribution has proven challenging in dc transport experiments due to the presence of often prevailing, conventional ABS's~\cite{Williams12,Sochnikov15,Kurter15}. Moreover, poisoning processes -- stochastic parity-changes of the quasiparticle occupation number -- may obscure the MBS contribution by limiting its lifetime~\cite{Rainis12,Albrecht17}. Measurement schemes probing at timescales shorter than this lifetime are thus essential. The Shapiro effect comes forth with the combined advantages of a radio-frequency ($f_{r\!f}$) excitation of the phase that can be faster than the poisoning dynamics~\cite{Fu09,Houzet13,Badiane13}, and the ease of dc current-voltage ($IV$'s) characteristics measurements. 

The immediate consequence of the fractional ac Josephson effect is an unusual sequence of Shapiro voltage steps $\Delta V =  \frac{h f_{r\!f}}{e}$ in the $IV$ characteristics, twice of that of conventional Shapiro steps ($\frac{h f_{r\!f}}{2e}$)~\cite{Kitaev01,Jiang11,Dominguez12,Houzet13,Badiane13,Virtanen13,Dominguez17}, providing direct evidence for the MBS $4\pi$-periodicity. First experiments performed on InSb nanowires~\cite{Rokhinson12}, on strained HgTe 3D TI~\cite{Wiedenmann16}, and on Bi$_{1-x}$Sb$_x$~\cite{Li17} junctions however showed surprises in the sequences of Shapiro steps. In all these cases, only the $n=\pm 1$ steps were absent in a given range of $r\!f$ power and frequency ($n$ is the integer index of the Shapiro steps), an absence which was described as an (incomplete) signature of the fractional a.c. Josephson effect. More recent measurements on 2D HgTe quantum wells showed the absence of odd steps up to $n=9$~\cite{Bocquillon17} and Josephson radiation at half the Josephson frequency~\cite{Deacon17}, though without the demonstration of time-reversal symmetry breaking that is required to induce MBS's in quantum spin-Hall edge channels~\cite{Fu09,Sticlet18}. While the latter observations advocate more clearly for the existence of $4\pi$-periodic Andreev modes, the fact that the  fractional ac Josephson effect acts only on some odd Shapiro steps depending on the system remains unclear.  Whether it provides a signature of the Majorana mode is a central question for identifying topological superconductivity in a variety of implementations.

In this work we report on the observation and understanding of the partial fractional ac Josephson effect in Josephson junctions designed on exfoliated flakes of the 3D topological insulator Bi$_2$Se$_3$. Our devices exhibit an anomalous sequence of Shapiro steps with (only) the first step absent at low $r\!f$ power and frequency. The $4\pi$-periodic contribution to the supercurrent is directly identified as a residual supercurrent at the first node of the critical current when the $r\!f$ power is increased. To shed light on our findings, we develop a two-channel Resistively Shunted Junction (RSJ) model that includes the quasiparticle overheating induced by Joule effect~\cite{Likharev79,Courtois08,DeCecco16}, and a thermally activated poisoning of the MBS. We show that Joule overheating suppresses the parity lifetime of the MBS and thus terminates the $4\pi$-periodic contribution to any higher index Shapiro steps, accounting for the observed suppression of the first Shapiro step only.

\vspace{0.5cm}

\textbf{Results}

\vspace{0.5cm}
%%%%%%%%%%%%%%%%%%%%%%%%%%%%%%%%%%%%%%%%%
\textbf{Partial even-odd effect in Bi$_2$Se$_3$ Josephson junctions.} 
Our samples are based on flakes of Bi$_2$Se$_3$ crystals exfoliated with the scotch tape technique. Figure~\ref{Fig1}c shows a $30 \, $nm thick  flake of  Bi$_2$Se$_3$ contacted with multiple electrodes of vanadium enabling both magneto-transport and Josephson junction measurements on the same  Bi$_2$Se$_3$ crystal (see Appendix for fabrication and measurement details). Analysis of Shubnikov-de-Haas oscillations and Hall effect enable identification of three electronic populations contributing to the sample conductance: Bulk states with a charge carrier density of $4.5 \times 10^{19}\,\text{cm}^{-3}$, and the top and bottom surface states with densities of $1\times 10^{12}$ and $4 \times 10^{12}\,\text{cm}^{-2}$ respectively (see Supplementary Note 1). All three channels may thus carry supercurrent by proximity effect~\cite{Sacepe11}.

We focus here our discussion on the Josephson junction of length $L=125\,$nm and width $W=2.25\,\mu$m (see geometry in Fig.~\ref{Fig1}b) indicated by the white arrow in Fig. \ref{Fig1}c.  Below the superconducting transition temperature of the electrodes ($T_c=5\,$K), the proximity effect develops in the TI, leading to a dissipationless supercurrent in the $IV$'s as shown in Figure~\ref{Fig1}d.  The transition to the resistive state of the junction ($R= 7.5\,\Omega$) is hysteretic at $0.05\,$K with switching and retrapping currents of $I_{sw} = 7.3\,\mu$A and $I_{r} = 5.0\,\mu$A, respectively. Such a hysteresis is a common feature of most Josephson junctions made with metallic weak links and results from a quasiparticle overheating in the normal section of the junction~\cite{Courtois08}. As we will show below, the ensuing quasiparticle overheating is key for understanding the suppression of the $n=\pm 1$ Shapiro steps only. 
\begin{figure}
	\includegraphics[width=0.9\linewidth]{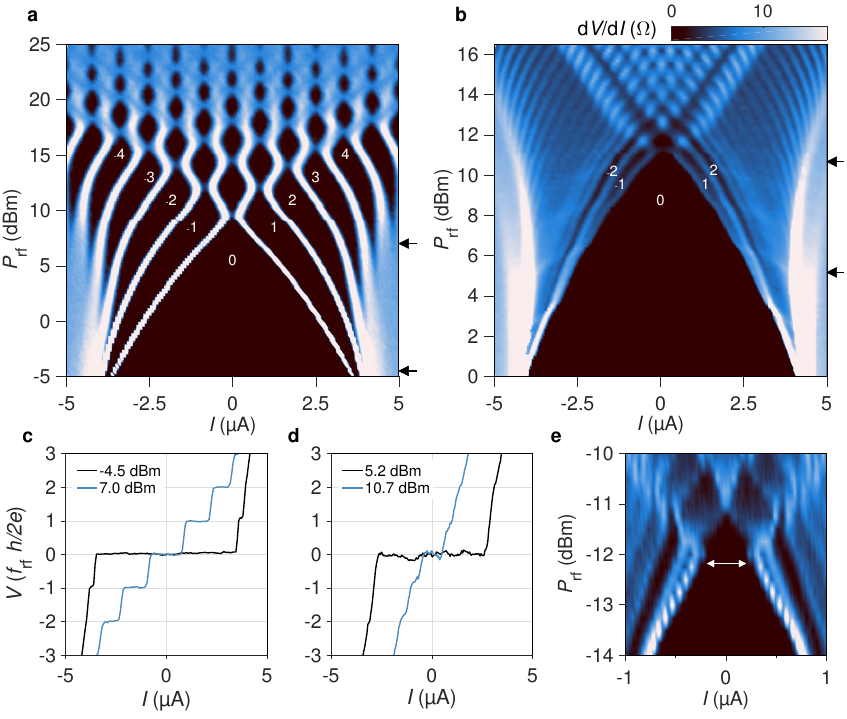} 
	\caption{\label{Fig2}\textbf{Partial fractional ac Josephson effect.} \textbf{a,b,} Shapiro maps displaying the differential resistance $dV/dI$ versus $r\!f$ power $P_{r\!f}$ applied to the antenna and dc current $I$ measured at $f_{r\!f}=3.5 \,$GHz in \textbf{a} and $ 1\,$GHz in \textbf{b}. White numbers indicate the Shapiro steps index.  \textbf{c,d,} Voltage $V$ in units of $f_{r\!f}h/2e$ versus $I$ extracted from the Shapiro map in \textbf{a} and \textbf{b} respectively at two different $P_{r\!f}$ values. The corresponding $P_{r\!f}$ values are indicated by the black arrows in \textbf{a} and \textbf{b}. \textbf{e,} Shapiro map displaying $dV/dI$ versus $P_{r\!f}$ and $I$, measured at a different frequency $f_{r\!f}=0.953\,$GHz during a second cooling of the sample, and zoomed on the first resistive node. The white arrow indicates twice the residual supercurrent ($2\times I_0^{k=1}$) at the first resistive node.}
\end{figure}

The dc response of the Josephson junction to an $r\!f$ irradiation is shown in the Shapiro map of Figure~\ref{Fig2}a that displays the color-coded differential resistance $dV/dI$ versus $r\!f$ power $P_{r\!f}$ and dc current $I$ for an $r\!f$ frequency of $3.5\,$GHz. At this frequency, well-defined Shapiro steps develop in the $IV$ curves, two of which are shown in Fig.~\ref{Fig2}c, with voltage steps that match the standard value $V_n =\pm n  \frac{h f_{r\!f}}{2e} = n\times 7.2\,\mu$V expected for a $2\pi$-periodic current-phase relation. In the Shapiro map, the black areas indicate $dV/dI=0$ and hence the position and amplitude of the Shapiro steps in the $P_{r\!f}-I$ plane. Two features standard for a usual Shapiro map are visible. First, on increasing $P_{r\!f}$, the critical current continuously decreases till nearly full suppression at $P_{r\!f}=9\,$dBm and then oscillates at higher $P_{r\!f}$. Second, the sequence of appearance of the Shapiro steps with $P_{r\!f}$ is sorted by the Shapiro step index $n$, and, importantly, starts with the step $n=1$.

The central experimental result of this work is displayed in Figure~\ref{Fig2}b, where we show the $r\!f$ response of the same junction at a lower frequency of $1\,$GHz.  This Shapiro map exhibits distinct features that markedly differentiate it from the higher frequency map. On increasing  $P_{r\!f}$, the first Shapiro step $n=1$ sets in at high $P_{r\!f}$ after the steps of higher index for both switching and retrapping currents. This anomaly, sometimes termed even-odd effect~\cite{Houzet13,Badiane13}, results in the conspicuous absence of the first Shapiro step in $IV$'s picked out at low $P_{r\!f}$, while steps of higher indexes already appears, see Fig.~\ref{Fig2}d. Our findings match those recently obtained on InSb nanowires~\cite{Rokhinson12}, strained HgTe 3D TI ~\cite{Wiedenmann16} and Bi$_{1-x}$Sb$_x$ alloy~\cite{Li17}, which were interpreted as a signature of a $4\pi$-periodic MBS contribution.

A second and new feature emerges at the first minimum of the critical current when the $r\!f$ power is increased, i.e. at $P_{r\!f}\simeq 10.7\,$dBm. Contrary to a conventional Shapiro map where a complete supercurrent suppression is expected at what can be called a resistive node, a small supercurrent $I_c\sim 190\,$nA remains. This appears clearly in the individual $IV$'s of Fig.~\ref{Fig2}d, see for instance the blue curve there. Figure~\ref{Fig2}e displays a similar Shapiro map obtained at a slightly different $f_{r\!f}$ in the low-frequency regime, but zoomed on the resistive node where the critical current is expected to vanish but does not. 
We shall see in the following that this residual supercurrent provides a direct signature of the presence of a $4\pi$-periodic mode in the junction.

\vspace{0.5cm}
%%%%%%%%%%%%%%%%%%%%%%%%%%%%%%%%%%%%%%%%%
\textbf{Determination of the coherent transport regimes.} 
Capturing the ABS spectrum of a Josephson junction on 3D TI's remains difficult as several conduction channels, including bulk and surfaces, intervene. Should all channels carry supercurrent, the nature of charge carriers in them may lead to virtually different regimes of coherent transport that we assess in the following. For bulk carriers, a rough estimate of the mean free path (see Supplementary Note 1) gives a Thouless energy $E_{th}=\hbar D/L^2 \simeq 417\,\mu$eV, smaller than the superconducting gap of the vanadium electrodes $\Delta = 800\,\mu$eV. This channel thus belong to the class of long diffusive Josephson junctions~\cite{Dubos01}. In contrast, for the topological surface states, the spin texture of the Dirac electrons stemming from the spin-momentum locking leads to a very strong scattering anisotropy which promotes forward scattering. As a result, the transport time $\tau_{tr}$ is expected to be significantly enhanced with respect to the elastic scattering time $\tau_e$, with ratio $\tau_{tr}/\tau_e $ up to $60$ depending on the disorder source~\cite{Culcer10}. Recent experiments on Bi$_2$Se$_3$ flakes combining field effect mobility and quantum oscillations assessed a ratio $\tau_{tr}/\tau_e \gtrsim 8$~\cite{Dufouleur16}. Taking the latter value as a conservative estimate and the surface state elastic mean-free path $l_e \simeq 28\,$nm of our sample (see Supplementary Note 1) leads to a transport length $l_{tr}\gtrsim 225\,$nm. These considerations suggest that surface transport is ballistic with, importantly, a non-zero probability for straight electronic trajectories impinging both electrodes. This is also consistent with signatures of ballistic transport over $300\,$nm evidenced in Bi$_2$Se$_3$ nanowires~\cite{Dufouleur13,Hong14}. 

Consequently, we consider the topological surface state channel as ballistic. As such, the relevant energy scale for the ABS's is $\hbar v_F /L = 2.8\,$meV with $v_F = 5.4\times 10^5\,$m/s the Fermi velocity~\cite{Xia09}. It is greater than $\Delta $, which shall lead to ABS's in the short ballistic limit. Theory then predicts that a $4\pi$-periodic MBS exists, even in the case where the Fermi level is far from the Dirac point of the surface states, which is here the experimentally relevant regime~\cite{Snelder13}. This MBS corresponds to ballistic trajectories impinging perpendicularly the superconducting electrodes, all other incidence angles yielding conventional $2\pi$-periodic ABS's~\cite{Fu08,Snelder13}.

Eventually, observability of the $4\pi$-periodicity theoretically implies a strong constraint on the Andreev spectrum: The MBS's must be decoupled from the quasiparticle continuum at $\phi=0$ and $2\pi$ to avoid direct transfer of quasiparticles into or from the continuum. Such a transition would indeed occur from the excited to the ground state every $2\pi$, restoring an effective $2\pi$-periodicity for the MBS. This detrimental effect can be remedied by adding a magnetic layer or magnetic field that break time reversal symmetry, and thus open a gap between the MBS and the quasiparticle continuum~\cite{Fu09,Jiang11,Badiane11,Houzet13,Snelder13}, as sketched in Fig.~\ref{Fig1}b. In our samples, the vanadium that we use as superconducting electrodes is known to form magnetic dopants in Bi$_2$Se$_3$ and eventually a ferromagnetic phase at large concentration~\cite{Zhang17}. Given that a smooth ion milling of the Bi$_2$Se$_3$ surface is processed before vanadium deposition, favoring vanadium diffusion into the Bi$_2$Se$_3$ crystal, there is presumably a magnetic layer or local magnetic moments at the superconducting interface as well as magnetic moments on the oxidized vanadium side surfaces of the electrodes. This singular configuration is likely to break time reversal symmetry on the scale of the junction, thus leading to the decoupling of the MBS from the continuum and to the ensuing observability of $4\pi$-periodicity in our Shapiro maps. 

\vspace{0.5cm}
%%%%%%%%%%%%%%%%%%%%%%%%%%%%%
%%%%%%%%%%%%%%%%%%%%%%%%%%%%%
\textbf{Two-channel RSJ model. }
To understand our experimental findings, we consider a RSJ model comprising a pure Josephson junction in parallel with a shunt resistor $R$. In the usual scheme of a single Josephson channel with a critical current $I_c$, the key parameter for the phase dynamics is the phase relaxation time $\tau_J = \frac{\hbar}{2eRI_c}$ which sets the typical time scale for the phase to adapt to a drive current change~\cite{Likharev79}. With a $r\!f$ drive, the RSJ model thus acts on the phase as a low pass filter of cutoff frequency $1/\tau_J$ (see Supplementary Note 3). The regime of visibility of Shapiro steps is thus defined by  $f_{r\!f} \tau_J<1$. 

\begin{figure}[h!]
	\includegraphics[width=1\linewidth]{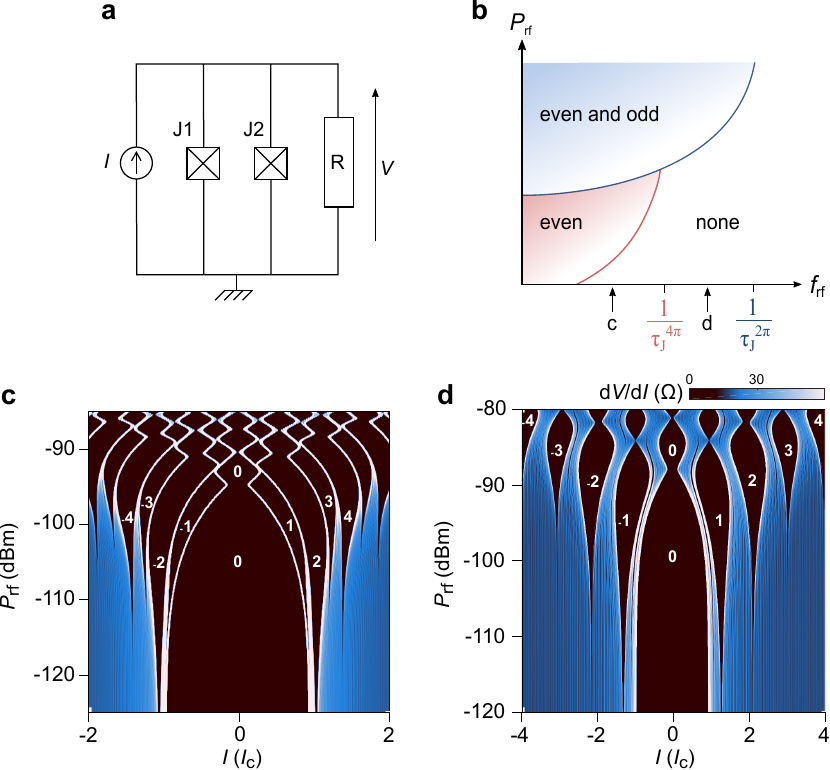} %,bb=0 0 192mm 175mm
	\caption{\textbf{Two channel RSJ model.} \textbf{a,} Equivalent circuit of the two-channel RSJ model comprising two Josephson junctions J1 and J2 of different critical current and current-phase periodicity, that is, $I_s = I_c^{2\pi}\sin(\phi)$ and $I_s = I_c^{4\pi}\sin(\phi/2)$ respectively, in parallel with the shunt resistor $R$. \textbf{b,} Range of existence of the even and odd Shapiro steps versus $f_{r\!f}$ and $P_{r\!f}$. $1/ \tau_J^{2\pi}$ and $1/ \tau_J^{4\pi}$ indicate the cutoff frequency above which the contribution to the Shapiro steps of their respective channel is strongly suppressed. \textbf{c, d,} Computed Shapiro maps for the two-channel RSJ model displaying the differential resistance $dV/dI$ versus normalized current $I/I_c$ and $P_{r\!f}$. The model parameters are presented in the Supplementary Note 3. The product $f_{r\!f}\tau_J^{4\pi}=0.4$ and $1.5$ in \textbf{c} and \textbf{d} respectively.}
\label{Fig3} 
\end{figure}

To phenomenologically capture the complex dynamics of a topological Josephson junction where a MBS lies within (a majority of) ABS's, we include two different Josephson junctions J1 and J2 in the RSJ model (see Fig. \ref{Fig3}a)~\cite{Dominguez12,Dominguez17}. The first junction J1 stands for the conventional ABS's with a $2\pi$-periodic current-phase relation, and the second one J2 represents the $4\pi$-periodic MBS's.
Having both $2\pi$ and $4\pi$-periodic contributions in the total supercurrent $I_s(\phi) = I_c^{2\pi}\sin(\phi) + I_c^{4\pi}\sin(\phi/2) $  drastically changes the Shapiro steps sequence. The dynamics is now ruled by two different phase relaxation times $\tau_J^{2\pi}=\frac{\hbar}{2eRI_c^{2\pi}} $ and $\tau_J^{4\pi}=\frac{\hbar}{eRI_c^{4\pi}} $ set by the respective critical current amplitudes $I_c^{2\pi}$ and $I_c^{4\pi}$.  Consequently, the $4\pi$-periodic contribution will impact the junction dynamics only for drive frequencies $f_{r\!f} <1/\tau_J^{4\pi}$. This can be straightforwardly seen in the Shapiro maps that we obtained by numerically solving the RSJ equation together with the Josephson relation $V=\frac{\hbar}{2e}<\frac{d\phi}{dt}>$. In the low frequency limit $f_{r\!f}  \tau_J^{2\pi} < f_{r\!f}  \tau_J^{4\pi} < 1$, all even Shapiro steps develop at $r\!f$ power $P_{r\!f}$ lower than their neighboring odd steps, leading to the following appearance sequence $|n|=\left\{2,1,4,3,6,5...\right\}$ on increasing $P_{r\!f}$ (see Fig.~\ref{Fig3}c). 
Lowering $f_{r\!f} $ would enhance this even-odd effect and result ultimately in a quasi-suppression of odd Shapiro steps. Conversely, when the drive frequency is faster than $1/\tau_J^{4\pi}$ but still lower than $1/\tau_J^{2\pi}$, that is, $f_{r\!f} \tau_J^{4\pi} > 1 > f_{r\!f} \tau_J^{2\pi}$, the $4\pi$-periodic contribution is suppressed, restoring the regular sequence of Shapiro steps appearance $|n|=\left\{1,2,3,4...\right\}$ on increasing $P_{r\!f}$, as shown in Fig. \ref{Fig3}d.  In a 2D space ($f_{r\!f}$, $P_{r\!f}$), the ranges of existence of the even and odd Shapiro steps are sketched in Fig. \ref{Fig3}b. Importantly, the even-odd effect is robust even if $I_c^{4\pi}$ sounds negligible compared to $I_c^{2\pi}$, since the even-odd effect will always be present at low enough frequency as soon as $f_{r\!f} < 1/\tau_J^{4\pi}$. This low-frequency observability of the even-odd effect furthermore excludes an explanation for the existence $4\pi$-periodic contribution based on Landau-Zener transitions at a soft gap, which should be otherwise enhanced at high frequencies.

\begin{figure}
	\includegraphics[width=1\linewidth]{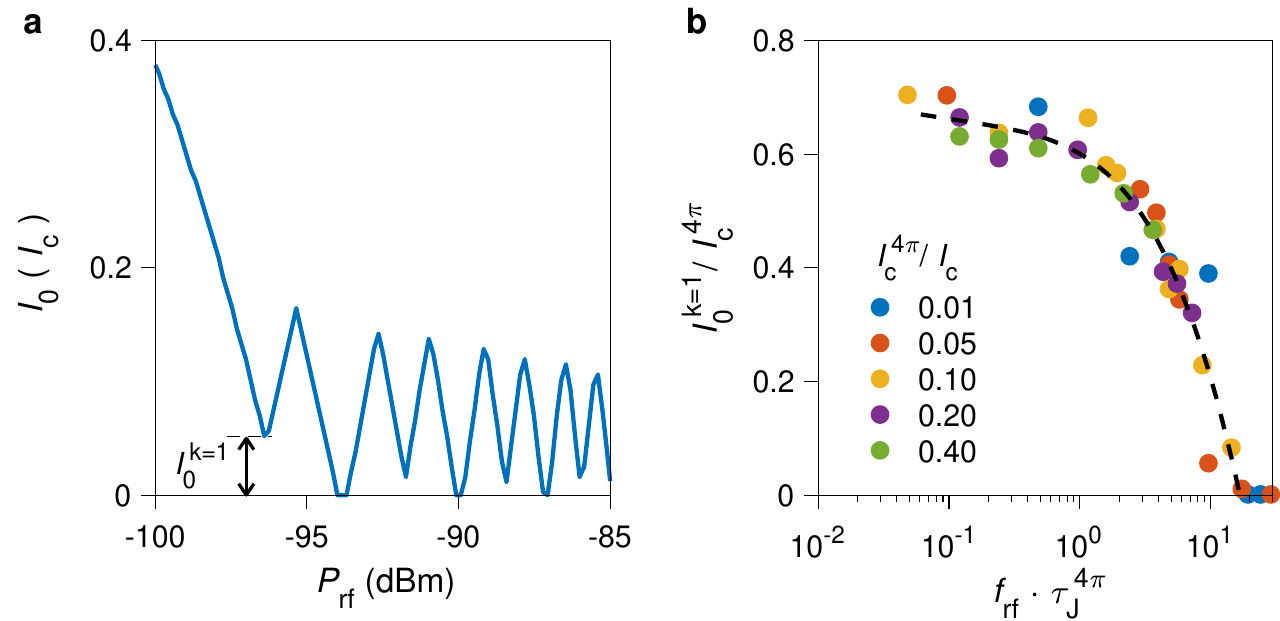} 
	\caption{\textbf{Residual supercurrent.} \textbf{a,} Switching current of the Shapiro step $n=0$ (supercurrent branch) $I_0$ normalized to the zero-power critical current $I_c$ versus ${r\!f}$ power $P_{r\!f}$. $I_0$ is extracted from Fig.~\ref{Fig3}c. The residual supercurrent that remains at the first node $k=1$ is noted $I_0^{k=1}$. \textbf{b,} Evolution of $I_0^{k=1}$ normalized to $I_c^{4\pi}$ versus $f_{r\!f}\,\tau_J^{4\pi}=f_{r\!f}\hbar / e R I_c^{4\pi}$ computed for various  $I_c^{4\pi}/I_c$ ratios. All points collapse on a single curve. The black dashed line is a polynomial fit that enables to extract $I_c^{4\pi}$ from experimental parameters. Note that fluctuations of $I_0^{k=1}/I_c^{4\pi}$ around this dashed line stem from the limited sampling of the  Shapiro maps computed.}
	\label{Fig6}
\end{figure}

Let us now consider in the same computed maps the $P_{r\!f}$-dependence of the Shapiro steps amplitude. In contrast with the standard oscillatory behavior with a complete suppression at the resistive nodes, the even steps, including the supercurrent branch ($n = 0$), exhibit a non-vanishing amplitude at every two resistive node on increasing $P_{r\!f}$, see Fig.~\ref{Fig3}c and d. This unusual feature has been predicted in a recent paper by Dom\'{\i}nguez \textit{et al.}~\cite{Dominguez17}. It is a direct consequence of the presence of the $4\pi$-periodic channel. It also conspicuously matches the residual supercurrent at the first node of the supercurrent branch in the experimental data of Fig.~\ref{Fig2}b and e.

We demonstrate below that the amplitude of this residual supercurrent can be \textit{quantitatively} related to the $4\pi$-periodic critical current $I_c^{4\pi}$. Figure~\ref{Fig6}a displays the computed switching current $I_0$ of the Shapiro step $n=0$, extracted from Fig.~\ref{Fig3}c as a function of $P_{r\!f}$. This plot both highlights the oscillatory behaviour of $I_0$ with $P_{r\!f}$ and enables us to identify the residual supercurrent of the first node that we note $I_0^{k=1}$, with $k$ the node index. To demonstrate the correlation between $I_0^{k=1}$ and $I_c^{4\pi}$, we performed a numerical study of the dependence of $I_0^{k=1}$ on the relevant parameters $I_c^{4\pi}$ and $f_{r\!f}$ by systematically computing Shapiro maps for different sets of parameters. Figure~\ref{Fig6}b displays the calculated $I_0^{k=1}/I_c^{4\pi}$ versus $f_{r\!f}\,\tau_J^{4\pi}$ for different  $I_c^{4\pi}/I_c$ ratios (we define here the total critical current $I_c = I_c^{2\pi} + I_c^{4\pi}$). All $I_0^{k=1}/I_c^{4\pi}$ values collapse on a single curve which tends to saturate to $\sim 0.7$ in the limit $f_{r\!f}\,\tau_J^{4\pi}\ll 1$. Conversely, $I_0^{k=1}$ vanishes when $f_{r\!f}\,\tau_J^{4\pi}\sim 16$, indicating that this residual supercurrent is visible at higher frequencies than the even-odd effect on the step appearance order. 

The remarkable consequence of the collapse is that $I_c^{4\pi}$ is uniquely defined for a fixed set of parameters $I_0^{k=1}$, $R$ and $f_{r\!f}$. Using a polynomial fit of the data points in Fig.~\ref{Fig6}b (black dashed line), we can thus directly determine $I_c^{4\pi}$ from the measured value of the residual supercurrent $I_0^{k=1}$ at a given frequency $f_{r\!f}$ (see Supplementary Note 5). With the experimental parameters of Fig.~\ref{Fig2}b, that is, $R=7.5\,\Omega$, $I_0^{k=1}=190\,$nA and $f_{r\!f}=1\,$GHz, we obtain $I_c^{4\pi}=290\,$nA and thus a ratio $I_c^{4\pi}/I_c\sim 4\%$. This value of $I_c^{4\pi}$ is close to the theoretical value of the supercurrent carried by a single mode $e\Delta / \hbar = 195\,$nA with $\Delta =0.8\,$meV being the superconducting gap of the vanadium electrodes~\cite{Beenakker91}, although the exact value of the induced gap is most likely smaller than $\Delta $ due to non-perfect interface transparency. Furthermore, the resulting values $f_{r\!f}\,\tau_J^{4\pi}\simeq 0.3$ and $1$ for the $1\,$GHz and $3.5\,$GHz Shapiro maps of Fig.~\ref{Fig2}b and a respectively are in agreement with the presence (absence) of even-odd effect in the data, confirming the consistency of the analysis. Note that our estimate of $I_c^{4\pi}$ is close those found on strained HgTe~\cite{Wiedenmann16} and Bi$_{1-x}$Sb$_x$~\cite{Li17} systems, which were based on the frequency cut-off criterion.  

\vspace{0.5cm}
%%%%%%%%%%%%%%%%%%%%%%%%%%%%%
%%%%%%%%%%%%%%%%%%%%%%%%%%%%%
\textbf{Joule-induced poisoning of the MBS. }
\begin{figure*}
	\includegraphics[width=0.8\linewidth]{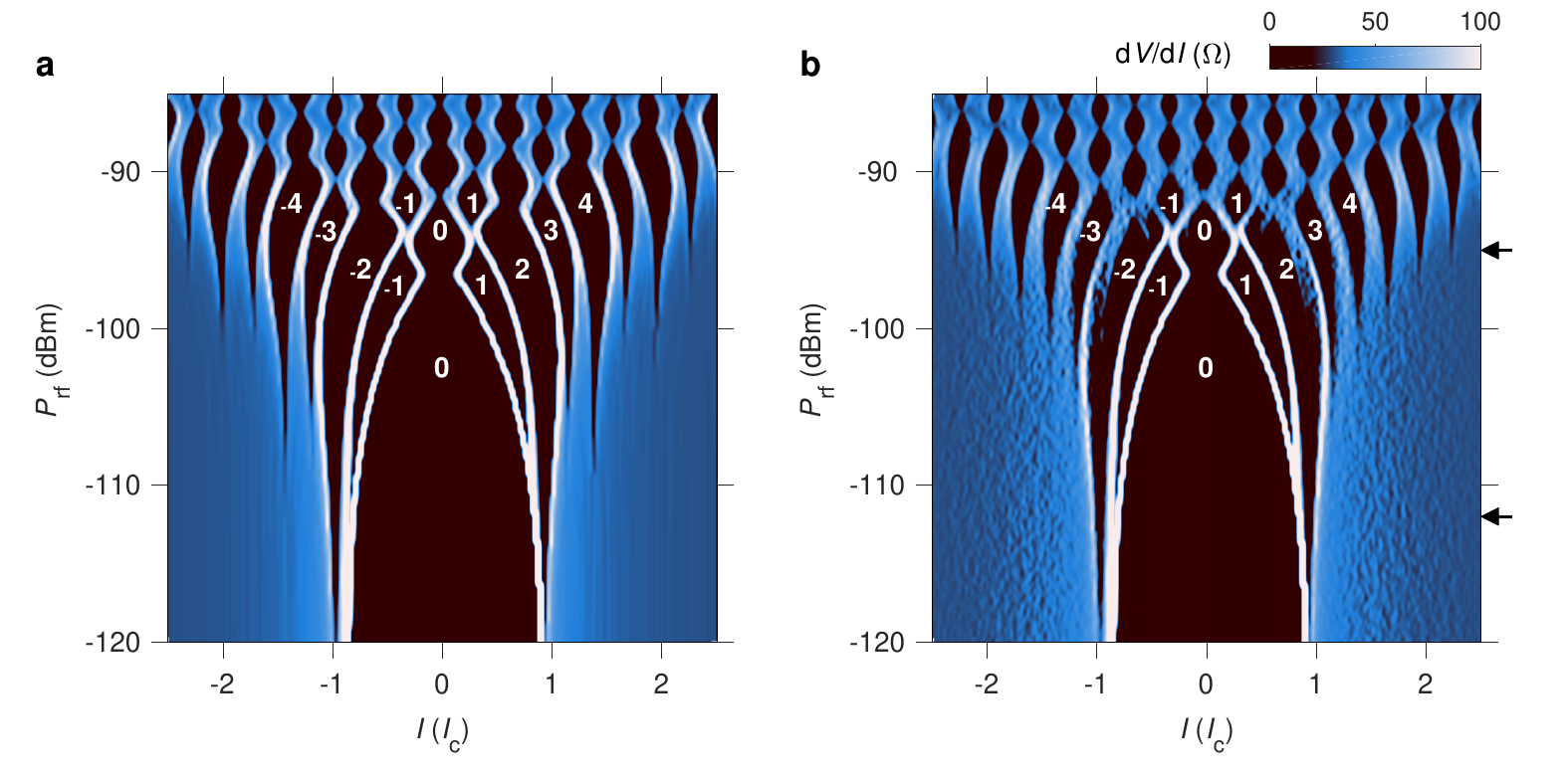} 
	\caption{\textbf{Poisoning of the even-odd effect.} \textbf{a,} Shapiro map computed with the two-channel thermal RSJ model displaying the differential resistance $dV/dI$ versus dc current $I$ normalized to $I_c$ and $P_{r\!f}$. For the sake of clarity, we took the same parameters as in Fig.~\ref{Fig3}, with a phonon bath temperature $T_{ph}=0.1\,$K. The model includes the experimental $T$-dependence for $I_c^{2\pi}(T)$ and an estimate of the electron-phonon coupling in Bi$_2$Se$_3$ (see Supplementary Note 6). On each point a recursive algorithm solves successively the RSJ equation, the ensuing time dependent voltage and dissipated power, the raise of the quasiparticle temperature $T_{qp}$, and then re-solves the RSJ equation with the new $I_c^{2\pi}(T_{qp})$ till convergence. \textbf{b,} Shapiro map computed with the two-channel thermal RSJ model as in \textbf{a} but including the thermal poisoning. The implementation of a thermally activated poisoning suppresses the even-odd effect for the Shapiro step of indexes $n\geq 3$.}
	\label{Fig4}
\end{figure*}
The two-channel RSJ model, however, does not account for the experimental absence of  the first Shapiro step only. Inclusion of a junction capacitance (geometric or instrinsic~\cite{Antonenko15}) through a two-channel RSCJ model mitigates the suppression of the $n\geq3$ odd steps~\cite{Pico17}, but is however not relevant for our strongly overdamped Bi$_2$Se$_3$-based Josephson junctions~\footnote{With a geometrical capacitance of the order of $10\,$aF estimated with a planar capacitor approximation for the superconducting electrodes, we obtain a damping parameter $\sigma =\sqrt{\hbar/2eI_cR^2C}\sim 260$. We also estimate an intrinsic capacitance $C^*\sim 0.1$~pF predicted in Ref.~\cite{Antonenko15} which leads to $\sigma = 2.5$. This overdamped regime is in contradiction with the hysteresis seen in the $IV$'s, consequently bearing out the electron overheating scenario.}. Furthermore, in the case of strongly underdamped, hysteresis should also be sizable in the Shapiro steps~\cite{Hamilton00}, which is not observed in our experiment. We propose instead that the absence of the first step can be explained by including thermal effects resulting from Joule heating, and their impact on the $4\pi$-periodic mode. Some of us recently showed that electron overheating must be taken into account with the RSJ model to capture Shapiro maps in conventional metallic Josephson junctions~\cite{DeCecco16}. Assuming that quasiparticles in the junction form a thermal distribution with an effective temperature $T_{qp}$ different from the phonon bath temperature $T_{ph}$,  we included the temperature dependence of the critical current, $I_c(T_{qp})$, and solved the RSJ equation self-consistently together with the heat balance equation $P=<I(t)V(t)>=\Sigma \Omega (T_{qp}^5 - T_{ph}^5)$ ($\Sigma $ is the electron-phonon coupling constant and $\Omega $ the volume of the normal part, see Supplementary Note 6 for values) to extract $T_{qp}$ for each dc current. Solving such a thermal RSJ model (tRSJ), we obtain a significant raise of $T_{qp}$, when a dc voltage drop sets in on the firstly developed Shapiro step~\cite{DeCecco16}.

We then conjecture that the ensuing excess of non-equilibrium quasiparticles that we express in terms of an effective quasiparticle temperature poisons the $4\pi$-periodic mode. Note that any non-thermal distribution would have the same consequences. Acting on a single mode, poisoning causes a stochastic parity-change~\cite{Fu09,Badiane11,Houzet13,Badiane13,Deacon17} that switches in time the quasiparticle occupation from the excited to the ground state, as illustrated in Fig.~\ref{Fig1}b by the black dotted arrow. Within the two-channel RSJ model, we model this parity-change of the $4\pi$-periodic contribution to the supercurrent as 
\begin{equation}
			I_s^{4\pi} (t)= (-1)^{n_{sw}(t)} I_c^{4\pi}\sin\left( \frac{\phi(t)}{2} \right),
	\label{eq:IcSwitch}
\end{equation}
where $n_{sw}(t)$ is a random occupation number of characteristic timescale determined by a switching parity lifetime $\tau_{sw}$ (see Supplementary Note 4).  The exact microscopic processes that yield such a diabatic event can involve several transitions, including pair breaking, quasiparticle recombination with various rates, and possible coupling to the bulk states. Phenomenologically, but without loosing the generality of the foregoing, we follow the approach of Fu and Kane~\cite{Fu09} and consider a thermally activated switching parity lifetime $\tau_{sw}$ for the $4\pi$-periodic mode:
\begin{equation}
\tau_{sw} = \tau_0 \exp\left( {\frac{\Delta - E_{MBS}}{k_B T_{qp}}} \right),
	\label{eq:tausw}
\end{equation}
where $\Delta - E_{MBS}$ is the minimal energy gap separating the $4\pi$-periodic mode to the continuum (see Fig.~\ref{Fig1}b), $\tau_0$ the characteristic timescale for the activation process, and $k_B$ the Boltzmann constant. When the quasi-particle effective temperature $T_{qp}$ is high, the switching time is small $\tau_{sw}\lesssim \tau_{J}$ and poisoning suppresses the $4\pi$-periodic Josephson effect, and hence the even-odd effect, by switching frequently the current-phase relation (see Supplementary Note 4). For the opposite limit $\tau_{sw} \gg \tau_{J}$, poisoning is irrelevant. 

We thus claim that understanding the partial even-odd effect in topological Josephson junctions relies on the interplay between activated poisoning and thermal effects in the two-channel tRSJ model. We show in Fig.~\ref{Fig4}a-b two Shapiro maps computed for the same $r\!f$ frequency and junction parameters as those in Fig. ~\ref{Fig3}c. Fig.~\ref{Fig4}a is the result of the tRSJ model that includes a $T$-dependence of $I_c^{2\pi}(T)$ given by the measured $I_{sw}(T)$~\cite{DeCecco16} and realistic parameters for the heat balance equation (see Supplementary Note 6 for an estimate of $\Sigma \Omega$ in our junctions). We assume here that most of the $T$-dependence of $I_{sw}(T)$ comes from the $2\pi$-periodic modes. Compared to Fig.~\ref{Fig3}c, the electron overheating leads to a broadening of the resistive transitions at high $P_{r\!f}$ between the periodic oscillations of the Shapiro steps.  Nevertheless, the full even-odd effect acting on all odd steps remains as in the absence of thermal effect, see Fig.~\ref{Fig3}c.

Figure~\ref{Fig4}b is the main result of our theoretical analysis. It displays a Shapiro map computed with the same tRSJ than Fig.~\ref{Fig4}a, but including thermally activated poisoning of the $4\pi$-periodic channel defined by Eq. (\ref{eq:IcSwitch}) and  (\ref{eq:tausw}) which ensues from Joule overheating. The $4\pi$-periodic channel now impacts only the steps $n=\pm 2$ by enhancing their amplitude, therefore inverting  the appearance order on increasing $P_{r\!f}$ between step $n=1$ and $2$.  The sequence of appearance of all higher order steps turns out to be regularized due to the suppression of the $4\pi$-periodic contribution by poisoning. This finding, that is, \textit{the even-odd effect limited to the first Shapiro step only}, is in full agreement with our experiment shown in Fig.~\ref{Fig2}b and with works on other systems~\cite{Rokhinson12,Wiedenmann16,Li17}. 

\begin{figure}
	\includegraphics[width=0.7\linewidth]{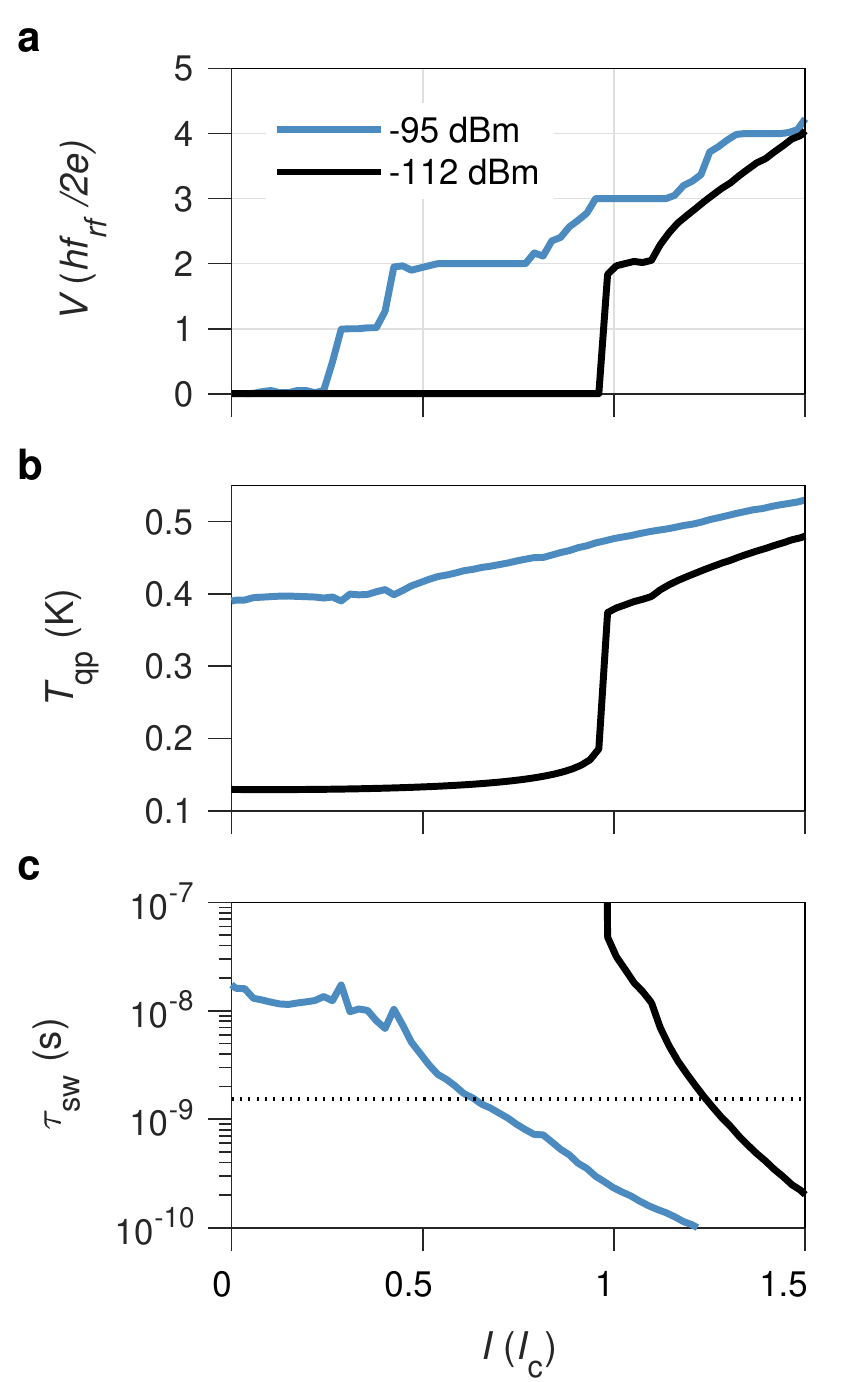} 
	\caption{\textbf{Thermally suppressed parity lifetime.} Data extracted from Fig.~\ref{Fig4}\textbf{b} at two different $P_{r\!f}$'s indicated by the black arrows in Fig.~\ref{Fig4}\textbf{b}, showing the voltage $V$ normalized to $hf_{r\!f}/2e$ (\textbf{a}),  $T_{qp}$ (\textbf{b}), and the switching time $\tau_{sw}$ (\textbf{c}) versus $I$ normalized to $I_c$. At $P_{r\!f}=-112$~dBm, the first Shapiro step is absent in the $I$--$V$. The increase of $T_{qp}$ when a voltage sets in on Shapiro steps leads to an exponential suppression of $\tau_{sw}$. When  $\tau_{sw}<\tau_J^{4\pi}$ ($\tau_J^{4\pi}$ is indicated by the black dotted line in \textbf{c}), poisoning suppresses the contribution of the $4\pi$-periodic channel to the Shapiro steps $n>2$, and all odd Shapiro steps are present in the $IV$, see blue curve in \textbf{a}.}
	\label{Fig4IV}
\end{figure}

The thermally activated poisoning can be captured by inspecting the computed $T_{qp}$ and $\tau_{sw}$ for two different $P_{r\!f}$'s. Figure~\ref{Fig4IV}a displays the $IV$ curves corresponding to the black arrows in  Fig.~\ref{Fig4}b. In Fig.~\ref{Fig4IV}b, we show the corresponding $T_{qp}$ versus $I$, which raises linearly once a dissipative voltage sets in. Accordingly, $\tau_{sw}$  is exponentially suppressed, and becomes inferior to $\tau_J$ on the Shapiro steps $n\geq2$ (Fig.~\ref{Fig4IV}c). This explains why the $4\pi$-periodic component acts only on the appearance order of the $n=1$ and $2$ steps, leaving all other steps unaffected. 

Furthermore, inspecting Fig. ~\ref{Fig4}b we see that the residual supercurrent of only the first resistive node $I_0^{k=1}$ remains in presence of poisoning, as observed in the data of Fig. \ref{Fig2}b. This confirms that the residual supercurrent  $I_0^{k=1}$ is a robust feature that provides a new indicator for the presence of $4\pi$-periodic modes, and enables a direct and quantitative determination of the corresponding $4\pi$-periodic critical current as discussed above. 

Within this approach of thermal poisoning by Joule overheating, it is interesting to compare the power dissipated between the available experiments on different TI systems. For instance, estimates of the dissipated power on the $n=2$ Shapiro step give two orders of magnitude difference between different TI materials: $P\simeq 2\frac{hf_{r\!f}}{2e}I\sim 25~\,$pW for our data and $17~\,$pW for strained HgTe~\cite{Wiedenmann16}. For InAs nanowires~\cite{Rokhinson12} and Bi$_{1-x}$Sb$_x$~\cite{Li17}, $P\sim 2~\,$pW. For HgTe quantum wells~\cite{Bocquillon17}, the dissipated power is the smallest $P\sim 0.2~\,$pW. Comparing the strained HgTe to the HgTe quantum wells that share the same electron-phonon coupling constant, we estimate a power per unit of volume of $\sim 80\,\text{fW}.\mu\text{m}^{-3}$ and $\sim 2\,\text{fW}.\mu\text{m}^{-3}$, respectively ($\sim 300\,\text{fW}.\mu\text{m}^{-3}$ in our Bi$_2$Se$_3$ samples). Such a small dissipated power density in the experiment on HgTe quantum wells should result in a minimized amount of non-equilibrium quasiparticles and limited poisoning, therefore explaining the observed full suppression of not only the first but of several odd Shapiro steps in this system.

\vspace{0.5cm}
%%%%%%%%%%%%%%%%%%%%%%%%%%%%%
%%%%%%%%%%%%%%%%%%%%%%%%%%%%%
\textbf{Discussion}

Our theoretical approach, combining two Josephson channels in parallel, electronic overheating and quasi-particle poisoning, can be extended to more elaborate situations. For instance, recent theory works predict for the case of the quantum spin Hall regime an $8\pi$-periodicicity due to either interactions~\cite{ZhangKane14} or quantum magnetic impurities~\cite{Peng16b,Hui17}. Although this has not been reported so far in experiments, it would be interesting to study how multiple periodicities mix in the Shapiro response. Given the understanding of the two-channel RSJ model, we expect the $8\pi$-periodicity to enhance every steps of index $\pm 4n$ and significantly modify the beating pattern at high $P_{r\!f}$ . Other effects such as the voltage dependence of the phase relaxation time~\cite{Houzet13} could be included in our model and should enhance the effect of thermal poisoning on the partial suppression of the even-odd effect.

To conclude, our work elucidates the origin of the puzzling suppression of only the first Shapiro step in topological Josephson junctions. In our Bi$_2$Se$_3$ Josephson junctions, this suppression is accompanied by a residual supercurrent that provides a new indicator of the $4\pi$-periodic contribution to the supercurrent. Together, these observations can be captured by a two-channel thermal RSJ model in which Joule overheating activates poisoning of the $4\pi$-periodic mode. The even-odd effect restricted to the first Shapiro step and the residual supercurrent do provide a clear signature of a $4\pi$-periodic mode in the Andreev spectrum, conspicuously pointing to MBS's. Our phenomenological model illustrates a direct consequence of thermal poisoning on Majorana bound states, signaling that dissipation must be scrutinized with attention in dc-biased measurements. Addressing a microscopic description of the enhanced poisoning in such non-equilibrium measurement schemes is a challenging task for theory that should lead to significant progress towards new devices for Majorana physics and possible MBS qubits.

\appendix* 
\section{Methods}

Bi$_2$Se$_3$ crystals were synthesized by melting growth method with high purity (5N) Bi and Se in an evacuated quartz tube. Crystals were analyzed by  X-rays diffraction, and angle-resolved photoemission spectroscopy. Flakes of Bi$_2$Se$_3$ were exfoliated from the bulk crystal on silicon wafer and systematically inspected by atomic force microscopy to ensure crystal quality. V/Au superconducting electrodes were patterned by e-beam lithography and deposited by e-gun evaporation after a soft ion beam etching. 

Measurements were performed in a dilution refrigerator equipped with highly filtered dc lines that comprise room temperature feed-through Pi-filters, lossy custom-made coaxial cables and capacitors to ground on the sample holder. Radio-frequency are fed through a dedicated coaxial cable ending as an antenna that were adjusted in the vicinity of the devices. Shapiro map measurements were performed with standard lockin amplifier techniques.

\vspace{1cm}

\textbf{Acknowledgments} We are grateful to Manuel Houzet and Julia Meyer for inspiring comments on the poisoning processes, and Teun Klapwijk for unvaluable discussions. We thank Jacques Marcus for his support on the crystal growth. Samples were prepared at the Nanofab platform and at the "Plateforme Technologique Amont" of Grenoble. This work was supported by the LANEF framework (ANR-10-LABX-51-01) and the H2020 ERC grant \textit{QUEST} No. 637815.
\vspace{1cm}

\bibliography{TJJ-bib}

%------------------------------------------------------
% Supplementary Information
%------------------------------------------------------

\clearpage
\onecolumngrid
\setcounter{figure}{0}
\setcounter{section}{0}
\renewcommand{\thefigure}{S\arabic{figure}}

%\doublespacing

%------------------------------------------------------
\begin{center}
\textbf{\large Supplemental Material for \\ Joule overheating poisons the fractional ac Josephson effect in topological Josephson junctions} \vspace{5mm}
\end{center}
%------------------------------------------------------

%%%%%%%%%%%%%%%%%%%%%%%%%%%%%%%%%
\section{I. Transport properties of the $Bi_2Se_3$ flakes}
\begin{figure}[h!]
	\includegraphics[width=0.7\columnwidth]{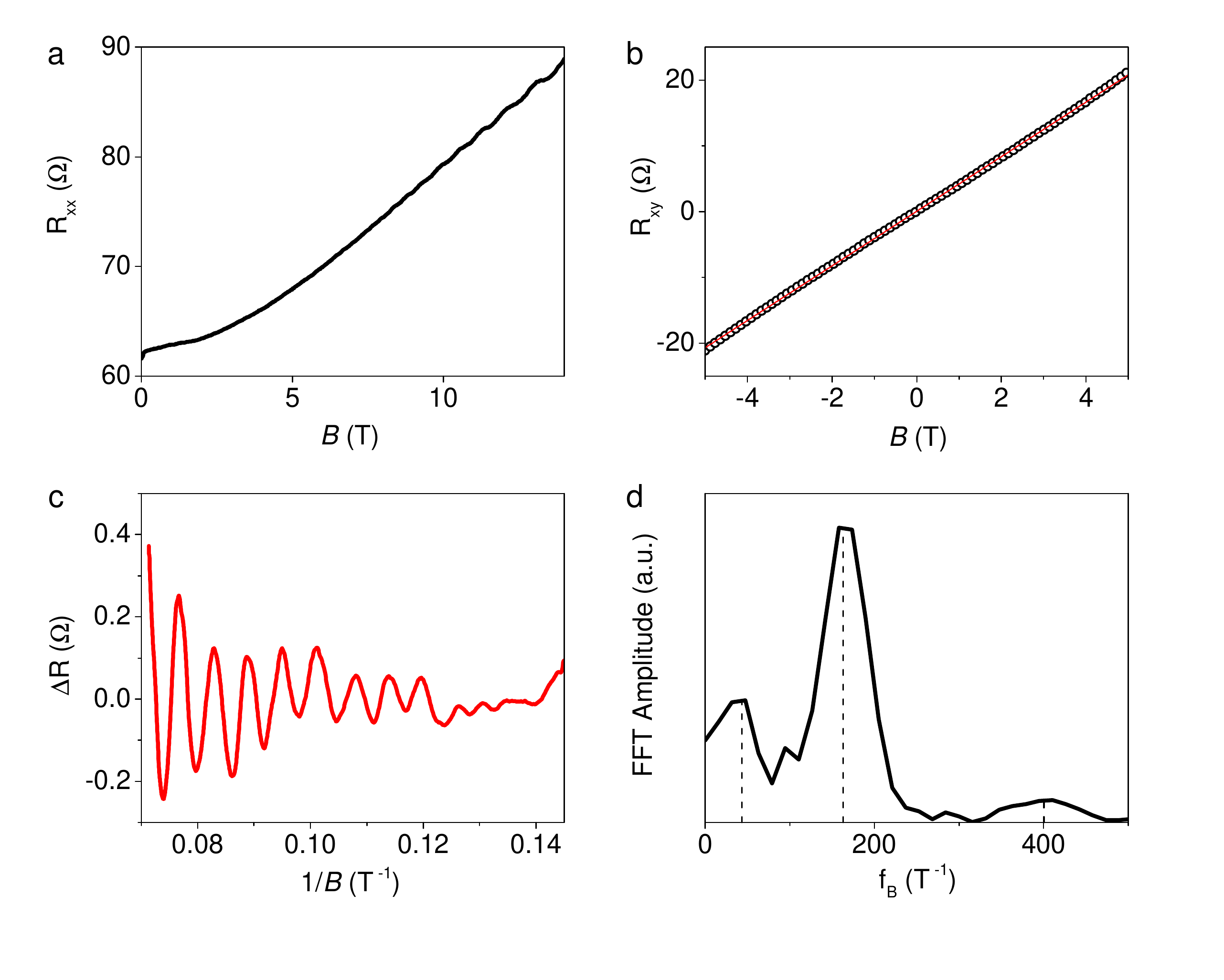} 
	\caption{\textbf{Magnetotransport of the Bi$_2$Se$_3$ flake.} \textbf{a,} Longitudinal resistance R$_{xx}$ as a function of magnetic field $B$. Shubnikov-de-Haas oscillations develop above $7\,$T. \textbf{b,} Hall effect in the Bi$_2$Se$_3$ flake. Transverse resistance R$_{xy}$ as a function of magnetic field $B$ measured at $4\,$K. \textbf{c,} Shubnikov-de-Haas oscillations after smooth background subtraction. \textbf{d,} Fourier transform of the oscillations. Three peaks are observed, indicated by vertical dashed lines.}
	\label{fig_transport}
\end{figure}

We present in this section the magnetotransport properties of sample LC106 discussed in the main text and determine the possible contributions to the conductance of the bulk and surface states.
The magnetoresistance measured at $4\,$K and up to $14\,$T is shown in Figure \ref{fig_transport}a. Above $7\,$T, Shubnikov-de-Haas (SdH) oscillations develop on top of a linear magnetoresistance characteristic of 3D topological insulators~\cite{Wang12,Yan13}. The oscillations after background removal are shown in Figure~\ref{fig_transport}c. The oscillatory pattern displays one obvious frequency, but also a beating signaling the presence of additional ones. A Fourier transform shown in Fig.~\ref{fig_transport}d reveals three main peaks, at magnetic frequencies $f_{B1} = 43\,\text{T}$, $f_{B2} = 163\,\text{T}$, $f_{B3} = 400\,\text{T}$ (see Fig. \ref{fig_transport}d), similar to what was observed in previous works \cite{Qu13, Veyrat15, Dufouleur16}. In Bi$_2$Se$_3$ nanostructures, three electronic populations are expected to contribute to electronic transport: One bulk band and two topological surface states of the lower and upper surfaces~\cite{Analytis10, Qu13, Veyrat15, Dufouleur16}, with respective charge carrier densities $n_b$, $n_{SS1}$ and $n_{SS2}$. It is worth noticing that for such strong charge carrier densities no downward band bending, and consequently no trivial charge accumulation 2DEG are expected~\cite{Veyrat15, King11, Analytis10}. The total carrier density $n_{tot}$ is therefore the sum of these three carrier densities.

To identify the magnetic frequencies to the different populations, we use the total carrier density extracted from Hall measurements. Figure~\ref{fig_transport}b displays the transverse resistance measured in sample LC106. A linear fit yields a total carrier density of $n_{\text{tot}}^{\text{Hall}} = 1.5 \times 10^{14}\, \text{cm}^{-2}$ indicating a high electronic doping. From the different possible scenarios for the SdH frequencies repartition, only one is compatible with the Hall data. By assuming either bulk or surface origin for the different magnetic frequencies, we can calculate a total carrier density $n_{\text{tot}}^{\text{SdH}} = n_{\text{b}}\times d + n_{\text{SS1}} + n_{\text{SS2}}$, with $d = 30$nm the flake thickness. Attributing either $f_{B1}$ or $f_{B2}$ to the bulk contribution (and therefore, the other two to TSS) would result in a total density of $n_{\text{tot}}^{\text{SdH}} = 1.8 \times 10^{13}\, \text{cm}^{-2}$ or $4.6 \times 10^{13} \,\text{cm}^{-2}$, which is much too low compared to the total carrier density $n_{\text{tot}}^{\text{Hall}}$. On the contrary, if we associate $f_{B3}$ with bulk carriers, we obtain a total density $n_{\text{tot}}^{\text{SdH}} = 1.4 \times 10^{14}\, \text{cm}^{-2} \simeq n_{\text{tot}}^{\text{Hall}}$. We therefore conclude that $f_{B3}$ is related to the bulk band, leading to a bulk carrier density of $n_{\text{b}} = 4.5 \times 10^{19}\,\text{cm}^{-3}$, while $f_{B1}$ and $f_{B2}$ correspond to the topological surface states with $n_{\text{SS1}} \simeq 1 \times 10^{12}\,\text{cm}^{-2}$ and $n_{\text{SS2}} \simeq 4 \times 10^{12}\,\text{cm}^{-2}$, respectively. Given that the peak at $f_{B2}$ gives the largest contribution to the SdH oscillations, we attribute $f_{B2}$ to the top surface and $f_{B1}$ to the bottom surface which has most likely a lower mobility than the top one due to its proximity to the SiO$_2$ substrate.

With the above charge carrier densities, we evaluate the number of transport modes for each conduction channel: $N = \frac{W k_F}{\pi} \simeq 130 $ and $250$ for the bottom and top surface states respectively. For the bulk band, one obtains  $N = \frac{dW k_F^2}{\pi^2} \simeq 9000 $.

Moreover, the onset of SdH oscillations provides an estimate of the electronic mobility. Throughout the whole field range, $f_{B2}$ (corresponding to a TSS) is the main peak of the FFT and the main contribution to the SdH. As a result, the surface mobility estimated from $\mu_{\text{SS}} \times B_{\text{onset}} = 1$ is $\mu_{\text{SS}} \simeq 1200\, \text{cm}^{2}/\text{V.s}$. For topological surface states, the corresponding mean free path is $l_e = v_F \, \mu_{\text{SS}} \,m^* /e = 28\, \text{nm}$, with $v_F = 5.4\times 10^5\,$m/s the Fermi velocity~\cite{Xia09} and $m^* = \hbar \sqrt{\pi n_{\text{SS2}}}/v_F = 0.076\, m_e$ the cyclotron mass at this electronic density ($m_e$ the electron mass). This value of mean free path is of the same order of magnitude as reported in previous works~\cite{Taskin12, Dufouleur16}.  As explained in the main text,  anisotropic scattering of the topological surface states favors forward scattering, leading to a transport length $l_{\text{tr}} \ge 8 \, l_e$~\cite{Dufouleur16, Culcer10}. Ballistic properties of the surface states may then be retained over more than $230\,$nm, larger than the $125\, $nm of the junction studied in the main text.

We can futhermore estimate the diffusion coefficient of the bulk states assuming that the bulk mean free path is of the order of the surface state mean free path~\cite{Dufouleur16}. This leads to $D=v_F^{bulk} l_e /3 \simeq 100\,cm^2/s$ with $v_F^{bulk} = \hbar (3\pi^2 n_b)^{1/3}/m^*$  and $m^* = 0.12 m_e$.  For our $125\,$nm long Josephson junction, the resulting Thouless energy of the bulk states is thus $E_{th}=\hbar D/L^2 \simeq 417\,\mu$eV.

%%%%%%%%%%%%%%%%%%%%%%%%%%%%%%%%%
\section{II. Partial even-odd effect on a second sample}
\begin{figure}[h!]
	\includegraphics[width = 0.7\columnwidth]{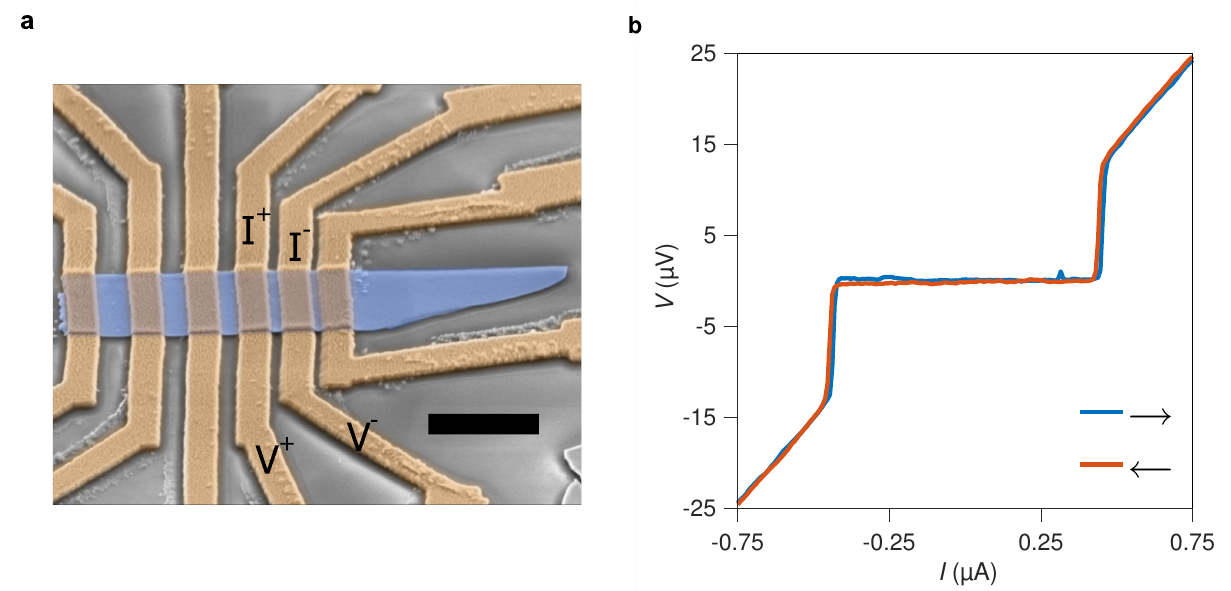} 
	\caption{\textbf{Josephson junction of a second sample.} \textbf{a,} Scanning electron micrograph of sample LC099 with false colors and tilted view. The  $20\,$nm thick flake of Bi$_2$Se$_3$ is in blue and the Ti/V/Au electrodes are in orange. The scale bar is $2\,\mu$m. The current ($I$) and voltage ($V$) contacts in black indicate the location of the junction. \textbf{b,} Current-voltage characteristics measured at $0.05\,$K.}
	\label{fig:LC099}
\end{figure}

We present in this section the Shapiro maps obtained on another sample that displayed a very similar behavior in terms of IVs characteristics and Shapiro maps. The Bi$_2$Se$_3$ flake is $20\,$nm thick and is contacted with Ti/V/Au ($5/70/5\,$nm) electrodes, see Fig.~\ref{fig:LC099}a. The dimensions of the Josephson junction are $W=1.45\,\mu$m and $L=140\,$nm. The normal state resistance is $R=30\,\Omega$ and the critical current shown in Fig.~\ref{fig:LC099}b is $490\,$nA. This value is smaller than the one of sample LC106 discussed in the main text. We explain this by the presence of the Ti layer beneath the vanadium which reduces the superconducting proximity effect. The absence of hysteresis at $0.08\,$K results from this small value of critical current which limits Joule heating so that no thermal bistability develops~\cite{Courtois08}. 

\begin{figure}[h!]
	\includegraphics[width = 0.9\columnwidth]{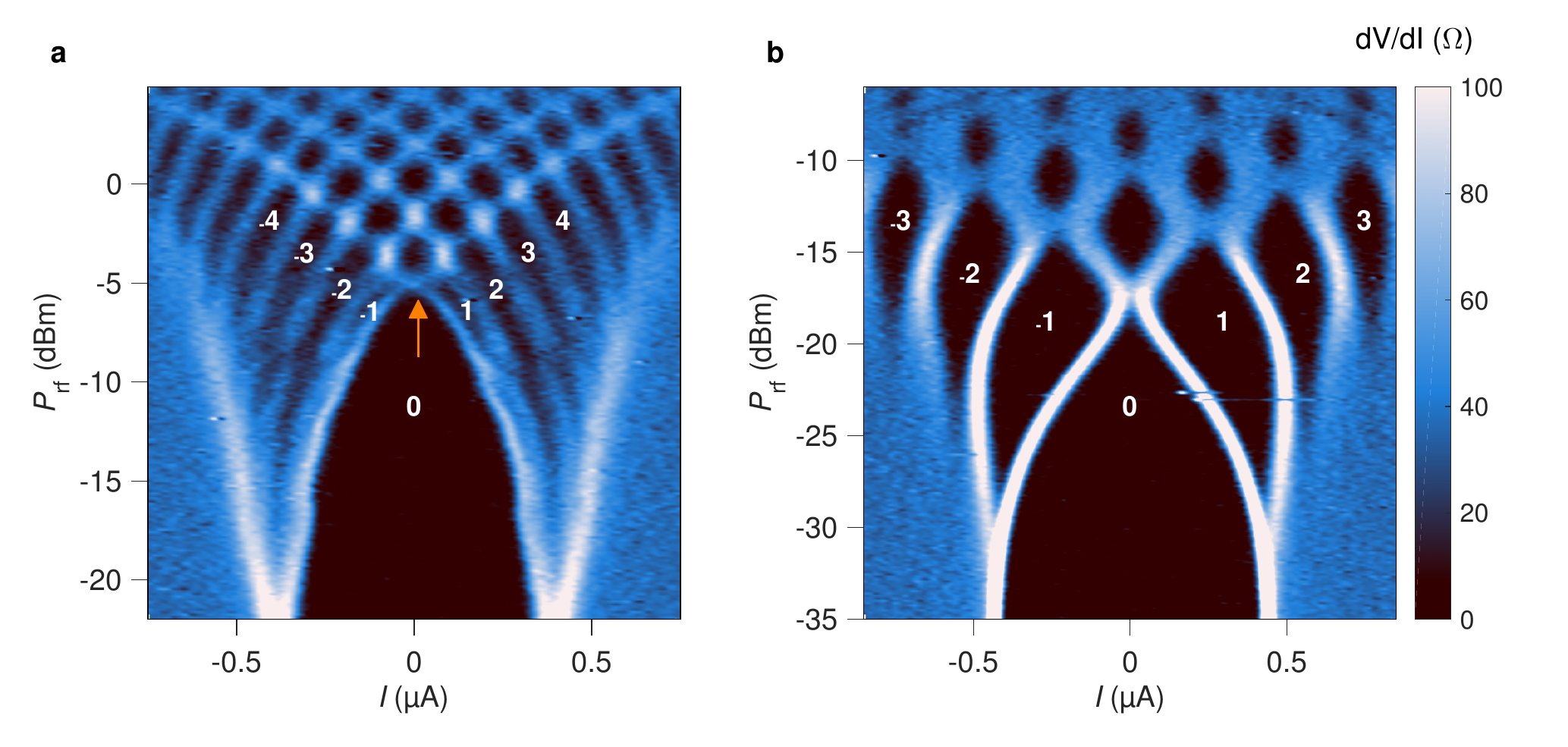} 
	\caption{\textbf{Shapiro maps of sample LC099.} Differential resistance $dV/dI$ versus current bias $I$ and $r\!f$ power $P_{r\!f}$ measured at $1.6\,$GHz (\textbf{a}) and  $4\,$GHz (\textbf{b}). The white numbers indicate the index $n$ of the Shapiro steps. Measurements were performed at $0.08\,$K. The orange arrow points to the first resistive node of the critical current, resistance value of which is mitigated compared to other nodes.}
	\label{fig:LC099Shapiro}
\end{figure}

Figure~\ref{fig:LC099Shapiro} displays two Shapiro maps of the junction measured at  $1.6\,$GHz (\textbf{a}) and  $4\,$GHz (\textbf{b}). Whereas the map at $4\,$GHz shows standard Shapiro steps, an even-odd effect develops in the map at lower frequency. The Shapiro step $n=1$ develops after the step $n=2$ on increasing $P_{r\!f}$, and has a smaller amplitude than the other steps. Furthermore, the resistance at the first node of the critical current, indicated in Fig.~\ref{fig:LC099Shapiro}a by the orange arrow, is less than at all other nodes. This is a reminiscence of the residual supercurrent observed in sample LC106 of the main text and consistent with the reduction of the first Shapiro step amplitude.

%%%%%%%%%%%%%%%%%%%%%%%%%%%%%%%%%
\section{III. Frequency response of the RSJ model}
\begin{figure}
	\includegraphics[width = \columnwidth]{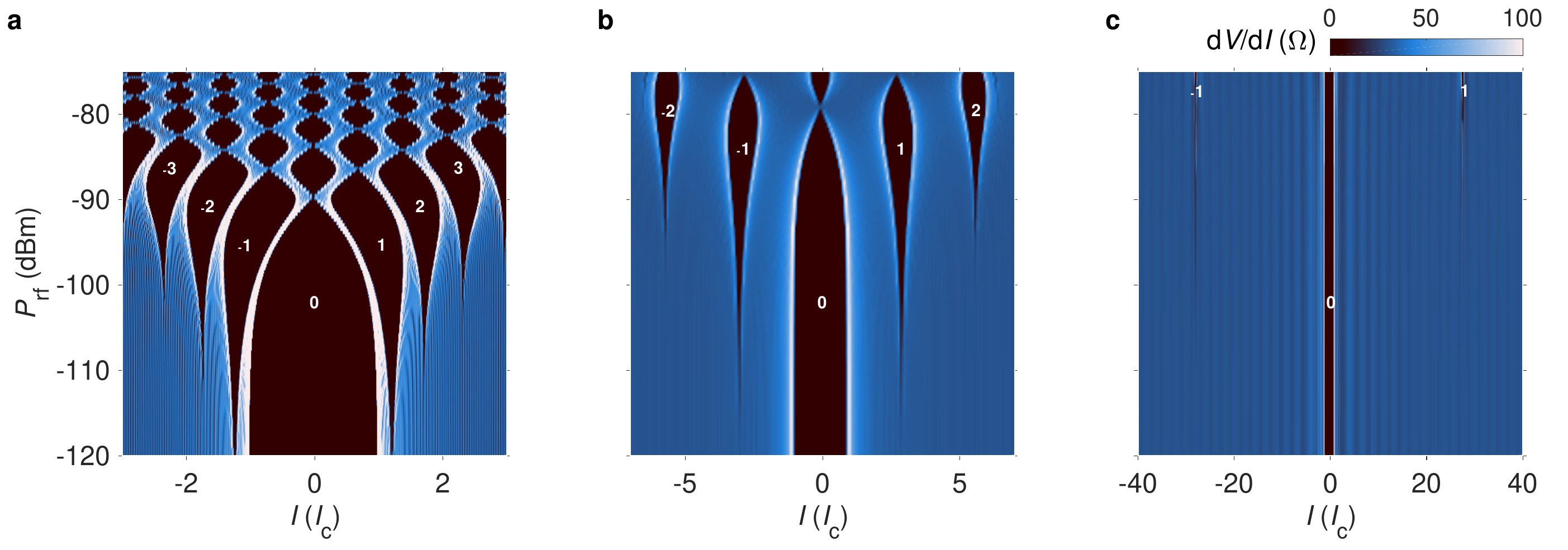} 
	\caption{\textbf{Frequency response of the RSJ model.} \textbf{a-c,} Shapiro maps displaying the differential resistance $dV/dI$ versus $I$ normalized to the critical current $I_c$ and $P_{r\!f}$, obtained by numerically solving a one-channel $2\pi$-periodic RSJ model. Parameters of the RSJ model are $R=30\,\Omega$, $I_c=500$~nA, and $ f_{r\!f}\tau_J^{4\pi}=0.22$, $0.88$ and $8.78$ ($f_{r\!f}=5$, $20$ and $200$~GHz) for \textbf{a}, \textbf{b} and \textbf{c} respectively.}
	\label{fig:FigS4}
\end{figure}

We present in this section the frequency response of the RSJ model. The RSJ equation: 
\begin{equation}
	\frac{d\phi}{dt} = \frac{2eR}{\hbar} \bigg[ I +  I_{r\!f} \sin (2\pi f_{r\!f}t) - I_s(\phi)   \bigg],
	\label{eq1:RSJ}
\end{equation}
features a characteristic phase relaxation time $\tau_J=\frac{\hbar}{2eRI_c}$ that acts as a low-pass frequency cutoff. For an $r\!f$ frequency faster than $1/\tau_J$, the phase cannot follow the $r\!f$  drive thus precluding the phase locking necessary to generate Shapiro steps. To illustrate this low-pass frequency response, we computed three Shapiro maps at different frequencies higher and lower than $1/\tau_J$. The resulting maps are displayed in Figure ~\ref{fig:FigS4}. For $f_{r\!f}\tau_J<<1$ in Fig. ~\ref{fig:FigS4}a, standard Shapiro steps develop, signaling that the phase is well driven by the $r\!f$ current, leading to phase-locking. For $f_{r\!f}\tau_J>>1$ in Fig.~\ref{fig:FigS4}c, the Shapiro steps are nearly absent. We distinguish only the $n=1$ step that starts to form at high $P_{r\!f}$ and develops a small current amplitude. This shows that the phase cannot follow the $r\!f$ drive above the cutoff frequency and a much higher power is needed to induce phase-locking. Therefore, our simulations shows that the RSJ model behaves as a low-pass filter.

%%%%%%%%%%%%%%%%%%%%%%%%%%%%%%%%%
\section{IV. Poisoning within the RSJ model}

We implement the parity-change of the $4\pi$-periodic channel by a stochastic sign--change of the current--phase relation:
\begin{equation}
	I_s^{4 \pi} (t) = (-1)^{n_{sw}(t)}  I_{c}^{4 \pi} \sin \left(\frac{\phi(t)}{2} \right),
			\label{eq:Ic4pipoisoning}
\end{equation}
where $n_{sw}(t)$ is a positive integer that counts the number of switching events while solving the time-dependent RSJ equation (\ref{eq1:RSJ}) for a given dc current bias $I$ and $r\!f$ drive. The value $n_{sw}$ at a time $t+dt$ where $dt$ is the time-step of the differential equation solver is given by:
\begin{equation}
	n_{sw}(t+dt) = n_{sw}(t) +  \mathcal{I} \left( x + 1 - \frac{dt}{\tau_{sw} }\right),
		\label{eq:nsw}
\end{equation}
where $x \in [0  \ 1]$ is a uniformly distributed random number, and $\mathcal{I}$ the rounding function to the nearest integer less than or equal to the value argument. For each time-step of the RSJ equation solving, the switching-probability is set by $ dt/\tau_{sw} $, which thus generates a Poisson-distributed parity-switching of mean lifetime $\tau_{sw}$.

\begin{figure}[h!]
	\includegraphics[width = 0.8\linewidth]{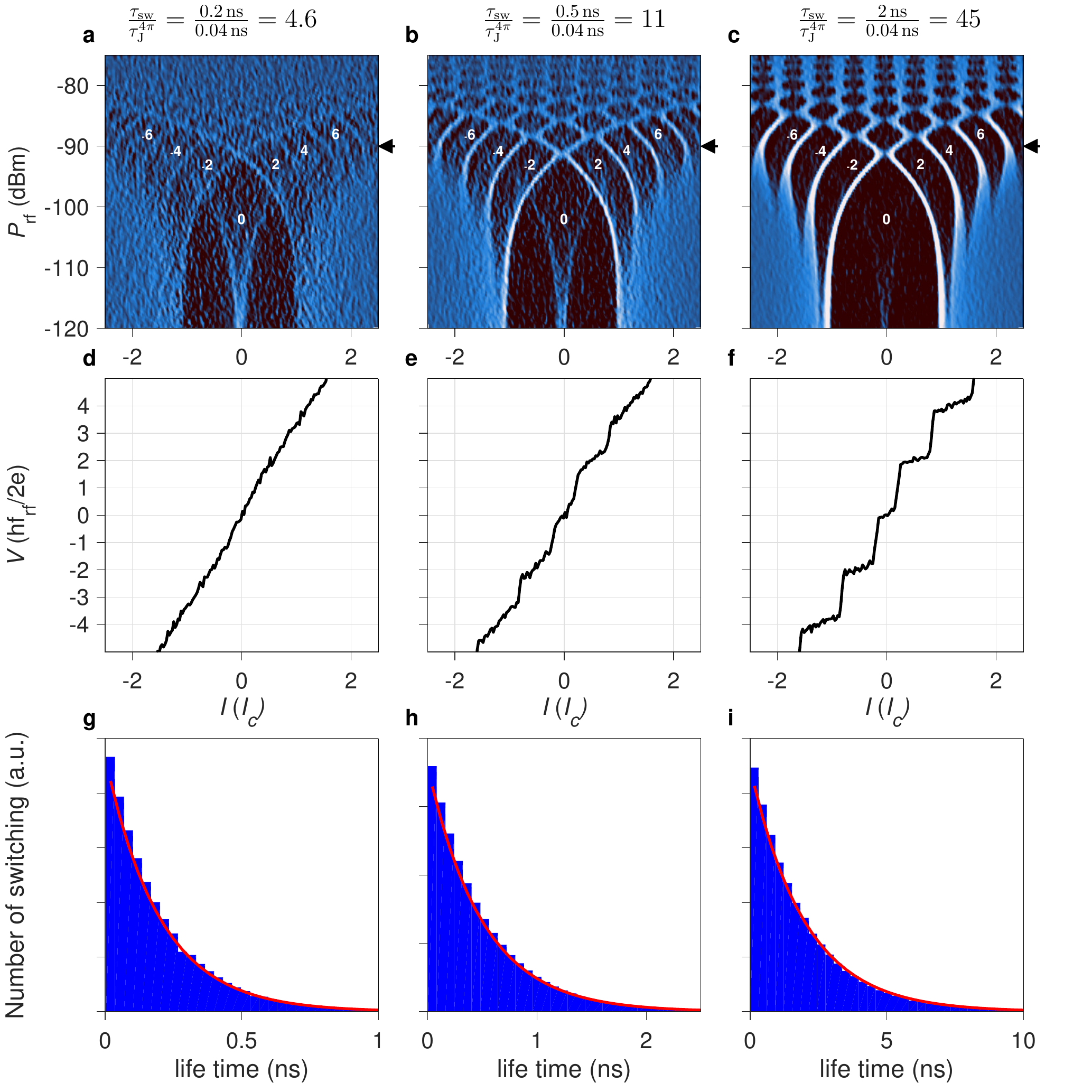} 
	\caption{\textbf{Impact of the switching time on the $4\pi$-periodic ABS.} \textbf{a-c,} Shapiro maps $dV/dI$ versus $I$ normalized to the critical current $I_c$ and $P_{r\!f}$, obtained by numerically solving a one-channel $4\pi$-periodic RSJ model subjected to poisoning. Parameters of the RSJ model are $R=30\,\Omega$, $I_c=500$~nA, and $f_{r\!f}=2$~GHz. The switching time is $\tau_{sw}=0.2$, $0.5$ and $2$~ns for (\textbf{a,d,g}), (\textbf{b,e,h}) and (\textbf{c,f,i}) respectively. \textbf{d-f,} $V$ in units of $hf_{r\!f}/2e$ versus $I$ normalized to $I_c$ extracted at the $P_{r\!f}$ indicated by the black arrows in \textbf{a-c}. 	For a large ratio $\tau_{sw}/\tau_{J}^{4\pi}$ in \textbf{a} and \textbf{d}, the Shapiro steps are barely visible. \textbf{g-i,} Histograms of the number of parity-switch versus parity lifetime for the respective Shapiro maps \textbf{a-c}. The red curves is a fit with an exponential decay which gives a mean value equal to $\tau_{sw}$.}
	\label{fig:Fig_tau_sw}
\end{figure}

Solving the RSJ equation (\ref{eq1:RSJ}) with (\ref{eq:Ic4pipoisoning}) enables to simulate the impact of poisoning on the Shapiro steps. Note that the current-phase relation is $4 \pi$-periodic in Eq. (\ref{eq:Ic4pipoisoning}). Figure \ref{fig:Fig_tau_sw}a-c displays three computed Shapiro maps obtained  for three different $\tau_{sw}$ values. We observe in Fig.~\ref{fig:Fig_tau_sw}a that when $\tau_{sw}$ is of the order of $\tau_J$, the poisoning suppresses the Shapiro steps and the $IV$'s are nearly ohmic (see the line-cut in Fig.~\ref{fig:Fig_tau_sw}d). 

 A Shapiro map computed in the opposite limit $\tau_{sw}\gg\tau_J$ is shown in Fig.~\ref{fig:Fig_tau_sw}c together with a line-cut in Fig.~\ref{fig:Fig_tau_sw}f. We see that a clear Shapiro steps develop, though with a small resistive slope due to the still non-zero poisoning. Consequently, when the mean switching time is much longer than the phase relaxation time, $\tau_{sw}\gg\tau_J$,  then poisoning becomes inefficient and lets the supercurrent and Shapiro steps nearly unaffected.

In Fig.~\ref{fig:Fig_tau_sw}g-i, we display histograms of the parity-lifetime for each  $\tau_{sw}/ \tau_J$ ratio. Exponential decay fits in red lines confirm the Poisson-distribution of mean $\tau_{sw}$. Note that the apparent noise Fig.~\ref{fig:Fig_tau_sw}a-f results from the finite time-span over which the RSJ equation is solved.

These simulations demonstrate how poisoning can affect and ultimately suppress supercurrent and Shapiro steps within the RSJ model. The phase relaxation time $ \tau_J$ is therefore the key parameter to define the sensitivity to the poisoning dynamics.

%%%%%%%%%%%%%%%%%%%%%%%%%%%%%%%%%
\section{V. Determination of $I_c^{4\pi}$ from the residual supercurrent}

\begin{figure}[h!]
	\includegraphics[width = 0.6\linewidth]{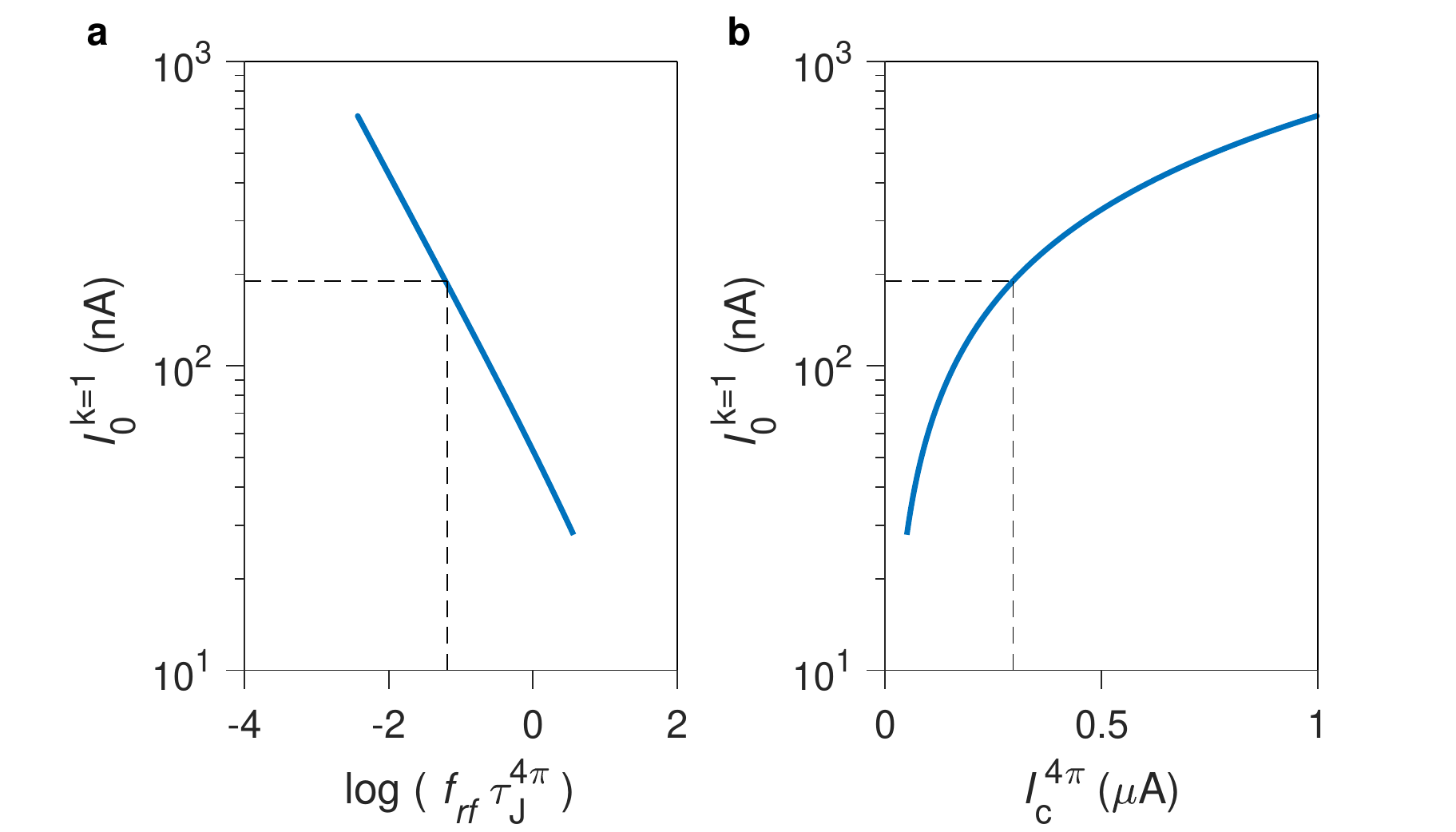} 
	\caption{\textbf{Determination of $I_c^{4\pi}$.} \textbf{a}. Residual supercurrent $I_0^{k=1}$ versus $\log \left( f_{r\!f}\,\tau_J^{4\pi} \right)$ computed for $R=7.5\,\Omega$ and $f_{r\!f}=1\,$GHz. \textbf{b}. $I_0^{k=1}$ versus $I_c^{4\pi}$. The black dotted lines indicate the experimental value of $I_0^{k=1}$ of Fig. 2b and the corresponding $\log \left( f_{r\!f}\,\tau_J^{4\pi} \right)$  and $I_c^{4\pi}$ values.}
	\label{fig:Fig_Ires}
\end{figure}

In this section we describe how to extract $I_c^{4\pi}$ from the experimental residual supercurrent and the scaling of $I_0^{k=1}/I_c^{4\pi}$ shown in Fig. 4b of the main text. The polynomial fit in Fig. 4b can be written $I_0^{k=1}/I_c^{4\pi} = \mathcal{P}\left( X \right)$ where $\mathcal{P}$ is a polynomial and $X = \log \left( f_{r\!f}\,\tau_J^{4\pi} \right)$. Using the experimental parameters of the junction discussed in the main text ($R=7.5\,\Omega$, $f_{r\!f}=1\,$GHz, $I_0^{k=1}=190\,$nA), we vary $I_c^{4\pi}$ and numerically compute $I_0^{k=1} = \mathcal{P}\left( X \right) I_c^{4\pi} $ versus $X$ (see Fig. \ref{fig:Fig_Ires}a) or versus $I_c^{4\pi}$ (see Fig. \ref{fig:Fig_Ires}b). For the data of Fig. 2b, we thus obtain $I_c^{4\pi} =  290\,$nA for $I_0^{k=1}=190\,$nA.

%%%%%%%%%%%%%%%%%%%%%%%%%%%%%%%%%
\section{VI. Electron-phonon coupling in Bi$_2$Se$_3$ Josephson junctions}
\label{sec:ElPh}

To solve the tRSJ model we estimated the electron-phonon coupling constant  $\Sigma_{Bi_2Se_3}$ of Bi$_2$Se$_3$ using the formula for dirty metals and two-dimensional phonons~\cite{Echternach92} : 
\begin{equation}
\Sigma  =   \left(\frac{k_B}{\hbar} \right)^5 \frac{4 \hbar \beta_l}{5 \pi^4} \; \frac{\Gamma(5) \zeta(5)}{v_F u_l^2} \; \frac{l_e}{d} \left(1+ \frac{3}{2} \left( \frac{u_l}{u_t}\right)^5 \right)
\end{equation}
with $u_{l,t}$ being the longitudinal and transverse sound velocities, $v_F$ the Fermi velocity $l_e$ the electronic mean free path, $d$ the film thickness, $\Gamma$ and $\zeta$ the gamma and zeta functions and $\beta_l$ a dimensionless parameter that we set to $1$~\cite{Echternach92}.  With $u_l \simeq 2900$~m/s and $u_t=1700$~m/s~\cite{Giraud12}, we estimate the electron-phonon coupling constant to be $ \Sigma_{Bi_2Se_3}  = 27.10^9~W.m^{-3} K^{-5}$.
\begin{figure}
	\includegraphics[width = 0.4\textwidth]{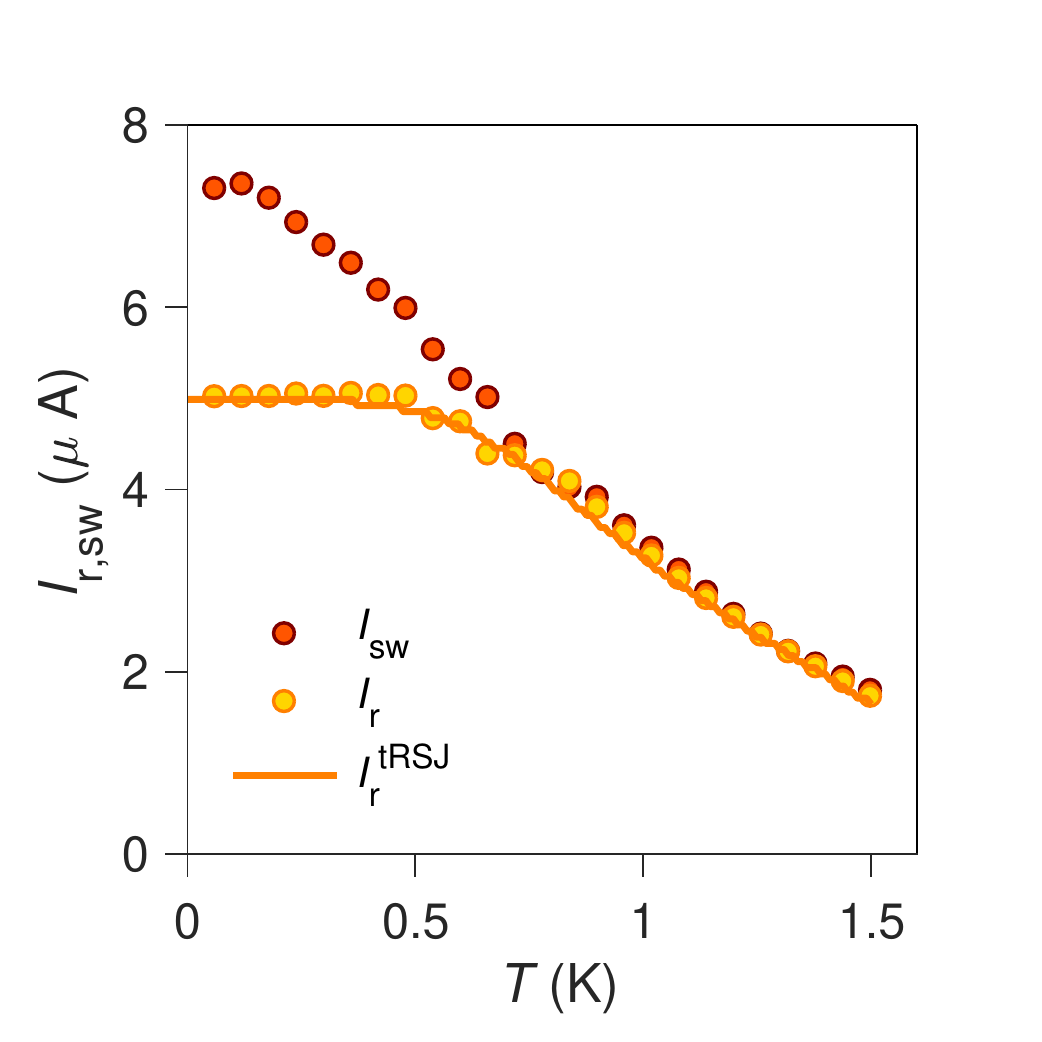} 
	\caption{\textbf{Retrapping current extracted from the tRSJ model.} The circles display the measured switching $I_{sw}$ and retrapping currents $I_{r}$ versus $T$ for the junction of the main text. The solid orange curve is the retrapping current computed with the tRSJ model with a volume $\Omega =  3.5.10^{-2}\,\mu m^3$ and $\Sigma_{Bi_2Se_3}  = 27.10^9~W.m^{-3} K^{-5}$.}
	\label{fig:LC106_Ic_vs_T}
\end{figure}

The validation of the above estimate can be done by fitting the $T$-dependence of the critical current hysteresis with the tRSJ model. Using the junction resistance, $ \Sigma_{Bi_2Se_3}$ and the measured switching current $I_{sw}(T)$ for each $T$ as input parameters, we adjusted the volume $ \Omega $ of the junction entering the heat-balance equation to fit the measured retrapping current $I_r$. Figure \ref{fig:LC106_Ic_vs_T} displays $I_{sw}$, $I_{r}$ and the resulting fit of $I_{r}$ for the junction studied in the main text. With the above value of $ \Sigma_{Bi_2Se_3}$, we obtain a volume equal to the product of the thickness $d$, width $W$ and the sum of the length $L$ and two halves of the electrode width ($400$~nm): $2.25\, \mu m\times 30\, nm \times (125\, nm+ 400\,nm) = 3.5 \times 10^{-2}\,\mu m^3$. The excellent agreement between fit and the data together with the resulting volume that is consistent with the junction geometry validate our estimation of $ \Sigma_{Bi_2Se_3} $. Overall, in our tRSJ model, the key parameter that controls the quasiparticle overheating is the product $\Omega \Sigma_{Bi_2Se_3} $, which we deduce from fitting the retrapping current.

%%%%%%%%%%%%%%%%%%%%%%%%%%%%%%%%%
\section{VII. Parameters of the RSJ model}
Solving the RSJ model with the parameters of the junction presented in the main text ($R=7.5\,\Omega$ and $I_c=7.3\,\mu$A) leads to a dense series of Shapiro steps, making the illustration of the two-channel RSJ model and the impact of thermal poisoning more delicate to capture visually. 

Therefore, for the sake of clarity, we choose in Fig.~3 and 5 parameters that provide large Shapiro voltage steps and well developed even-odd effect, namely, $I_c = 500$~nA, $R=30\,\Omega$, and $I_c^{4\pi}/I_c = 0.2$, and $f_{r\!f}=2\,$GHz. For Fig.~3d, $f_{r\!f}=7\,$GHz. 

Parameters for the thermal RSJ model in Fig.~5 are $ \Sigma_{Bi_2Se_3}  = 27.10^9~W.m^{-3} K^{-5}$,  $\Omega =  3.5.10^{-2}\,\mu m^3$ (see section~VI). For the activated poisoning, given that the exact Andreev spectrum is unknown, we  conjecture that $E_{MBS}\ll \Delta$ and thus parametrize $\Delta -E_{MBS} = 800\,\mu$eV~\cite{Fu09}. We thus empirically choose $\tau_0=0.8\times 10^{-19}$~s such that $\tau_{sw}$  drops with the computed $T_{qp}$ of the Shapiro map below $\tau_J^{4\pi}$. Importantly, other choice of $\Delta -E_{MBS} $ with $E_{MBS} $ closer to the continuum would lead to virtually the same effects.

%\bibliography{TJJ-bib}

\end{document}